# Monte Carlo Ray-Trace Diffraction Based On the Huygens-Fresnel Principle

J. R. Mahan,[1] N. Q. Vinh,[2,*] V. X. Ho,[2] N. B. Munir[1]

[1]Department of Mechanical Engineering, Virginia Tech, Blacksburg, VA 24061, USA
[2]Department of Physics and Center for Soft Matter and Biological Physics, Virginia Tech, Blacksburg, VA 24061, USA
*Corresponding author: vinh@vt.edu

**The goal of this effort is to establish the conditions and limits under which the Huygens-Fresnel principle accurately describes diffraction in the Monte Carlo ray-trace environment. This goal is achieved by systematic intercomparison of dedicated experimental, theoretical, and numerical results. We evaluate the success of the Huygens-Fresnel principle by predicting and carefully measuring the diffraction fringes produced by both single slit and circular apertures. We then compare the results from the analytical and numerical approaches with each other and with dedicated experimental results. We conclude that use of the MCRT method to accurately describe diffraction requires that careful attention be paid to the interplay among the number of aperture points, the number of rays traced per aperture point, and the number of bins on the screen. This conclusion is supported by standard statistical analysis, including the adjusted coefficient of determination, $R_{\text{adj}}^2$, the root-mean-square deviation, RMSD, and the reduced chi-square statistics, $\chi_v^2$.**

## 1. INTRODUCTION

The Monte Carlo ray-trace (MCRT) method has long been utilized to model the performance of optical systems in the absence of diffraction and polarization effects.[1-10] Heinisch and Chou,[11] and later Likeness,[9] were among the early proponents of treating diffraction in the MCRT environment. However, their approach, which is based on a geometrical interpretation of the Heisenberg uncertainty principle, relies on empiricism to obtain adequate agreement with theory.[12, 13] More recently the Huygens-Fresnel principle[14] has been implemented to describe diffraction and refraction effects in the MCRT [15-18] and wavefront tracing [10] environments. The goal of the current effort is to establish the conditions and limits under which the Huygens-Fresnel principle accurately describes diffraction in the MCRT environment. The method can be applied in the whole range of electromagnetic wave including the infrared region. This goal is achieved by systematic intercomparison of dedicated experimental, theoretical, and numerical results supported by statistical analysis.

## 2. APPROACH

We evaluate the success of the Huygens-Fresnel principle in describing diffraction in the MCRT environment by comparing predicted diffraction fringes with experimentally observed fringes produced for various aperture-to-screen distances, for both single slits and circular apertures. Predictions are based on an analytical approach widely available in the literature, and on the MCRT method described here. We compare the results from the analytical and numerical approaches with each other and with the dedicated experimental results. Standard statistical analysis is used to characterize differences observed among the theoretical, numerical, and the experimental results.

## 3. EXPERIMENTAL APPRATUS AND PROCEDURE

Figure 1 is a schematic diagram of the apparatus used to obtain the experimental results reported here.

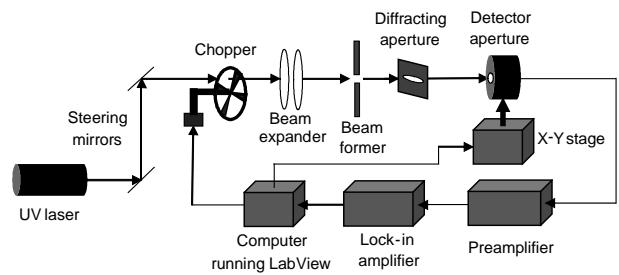

Figure 1. Schematic diagram of the experimental apparatus.

A 351-nm laser beam produced by a Coherent Enterprise II 653 Argon Laser System is steered through mirrors to a chopper. The chopper is used to modulate the intensity of the laser beam. After that it passes through a beam expander and a 4-by-4 mm beam former before falling on the aperture. The beam expander consists of two convex lenses whose focal lengths are 3.5 and 15 cm. The relatively large dimensions of the beam former ensure that the center of the beam does not contain a significant amount of diffracted light. The aperture consists of either a precision slit or a circular hole. The diffracted beam is incident to a 2.0-µm pinhole mounted on the entrance aperture of a Newport 918-UV photodetector. This pinhole determines the spatial resolution of the fringe measurements. Low-noise operation is assured by passing the detector output successively through a preamplifier and a lock-in amplifier. The lock-in amplifier is used to improve the signal-to-noise ratio of the setup. The shape and size of apertures and the aperture-to-pinhole spacing, z, are parameters of the study.

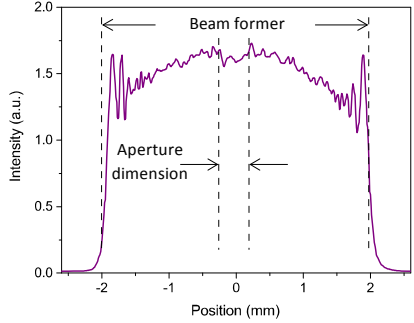

Figure 2. Scan across the expanded and formed laser beam cross-section between the beam former and the aperture.

Figure 2 shows an intensity profile, in arbitrary units, across the center of the expanded and formed laser beam cross-section immediately in front of the aperture. The vertical dashed lines indicate the maximum width of slits and apertures used. The intensity of the laser beam shows an accuracy of better than 2.5%. This image shows that the beam profile incident to the aperture is essentially flat to within the inherent noise level. The diffraction effects clearly visible at the edges of the beam former do not persist to the center of the beam.

## 4. ANALYTICAL DESCRIPTION OF DIFFRACTION

The diffraction irradiation pattern depends on the distance from an aperture to the observation screen. When the distance between the aperture and the observation is smaller than a wavelength, which is the near-field sub-wavelength region, the irradiation pattern has a shape similar to that of the aperture. When the irradiance pattern observed at a very great distance from the aperture ($z > a^2/\lambda$ where $a$ is the size of the aperture and $\lambda$ is the wavelength), we obtain the far-field pattern typical of Fraunhofer diffraction. The region in between the near-field sub-wavelength region and the far-field region is the Fresnel regime, or the near-field Fresnel region. We employ here an analytical description of diffraction for near-field Fresnel and the Fraunhofer diffraction.

Diffraction is considered to be in the Fresnel regime when either the light source or the observing screen, or both, are sufficiently near the aperture that the curvature of the wavefront becomes significant. Thus, we are not dealing with plane waves. Consider an aperture at $z = 0$ in the x', y'-plane illuminated with a monochromatic light of wavelength $\lambda$ and producing a field distribution, $E_0(x', y')$, within the aperture, as illustrated in Fig. 3. The field for the point P in the plane of observation (x, y), parallel to the x',y'-plane but at a distance z to the right, is given by adding together spherical waves emitted from each point in the aperture,[14, 19, 20]

$$E(P) = \frac{1}{i\lambda} \iint_\Sigma E_0(x', y') \frac{\exp(ikr)}{r} \cos\vartheta \, dx'dy'. \quad (1)$$

In Eq. (1), $\vartheta$ is an angle between a vector perpendicular to the x,y-plane and the vector $\vec{r}$ joining P and P'; thus $\cos\vartheta = z/r$. The distance between the points P and P' is given by

$$r = \sqrt{z^2 + (x - x')^2 + (y - y')^2}$$
$$\approx z\left(1 + \frac{1}{2}\left(\frac{x-x'}{z}\right)^2 + \frac{1}{2}\left(\frac{y-y'}{z}\right)^2\right). \quad (2)$$

Thus, we obtain the Fresnel approximation (near field),

$$E(x,y) = \frac{1}{i\lambda z} e^{ikz} e^{i\frac{k}{2z}(x^2+y^2)} \times$$
$$\iint_\Sigma E_0(x', y') e^{i\frac{k}{2z}(x'^2+y'^2)} e^{-i\frac{k}{z}(xx'+yy')} dx'dy'. \quad (3)$$

When both the source and the observation point are situated sufficiently far from the aperture (i.e., $z \gg k(x'^2 + y'^2)/2$), the factor $e^{i\frac{k}{2z}(x'^2+y'^2)}$ can be dropped from Eq. (3), yielding the Fraunhofer approximation (far field),

$$E(x,y) = \frac{1}{i\lambda z} e^{ikz} e^{i\frac{k}{2z}(x^2+y^2)} \times$$
$$\iint_\Sigma E_0(x', y') e^{-i\frac{k}{z}(xx'+yy')} dx'dy'. \quad (4)$$

We employ the Fourier transform to find the solution for both approximations.

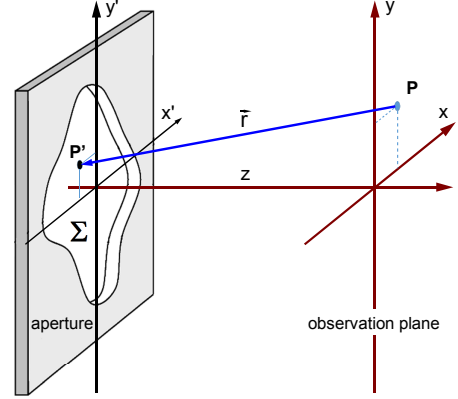

Figure 3. Transmission through an aperture.

## 5. MCRT DIFFRACTION MODEL

According to the Huygens-Fresnel principle, light propagates as a succession of self-replicating wavefronts. At its origin the wavefront for a plane wave is considered to consist of an array of equally spaced point disturbances, represented by the solid dots in Fig. 4a. Each point disturbance produces an outward-propagating pattern of concentric spherical waves. Then, for a given order of each spherical wavefront (t + Δt), a tangent plane is passed parallel to the original plane wavefront, with each point of tangency now considered to be a new point disturbance. As pointed out by Volpe, Létourneau, and Zhao,[10] "the construction should be regarded as a mathematical abstraction that correctly reproduces the physics without necessarily being physically rigorous."

It is convenient to recognize the duality between rays and waves in which the former are defined such that they are mutually orthogonal with the latter at points of intersection. The ray view of the Huygens-Fresnel principle is illustrated in Fig. 4b. In this view, each ray is considered to be an entity such as the one illustrated in Fig. 5; that is, it originates at a specified point, travels in a specified direction, and carries an electric field whose value varies periodically with position along its length as determined by the wavelength of the light.

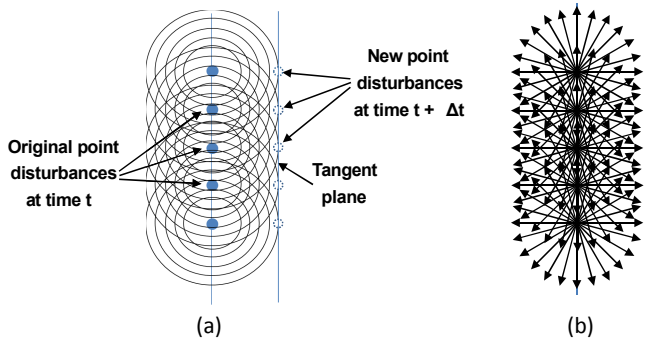

Figure 4. (a) Wave and (b) ray views of the Huygens-Fresnel principle for a right-running wave.

It is natural to identify the slit or circular aperture considered in this contribution as a plane source of rays of the type illustrated in Figs. 4b and 5. According to the Huygens-Fresnel principle, these rays will propagate from each source point in the slit or circular aperture with a directional distribution influenced by an obliquity factor. In the Monte Carlo ray-trace view of optics, source points randomly distributed in the plane of the slit or circular aperture emit rays with a directional distribution determined by an appropriate obliquity rule. Consistent with the assumption of a monochromatic plane wave incident on axis to the slit or aperture, all diffracted rays will be in phase.

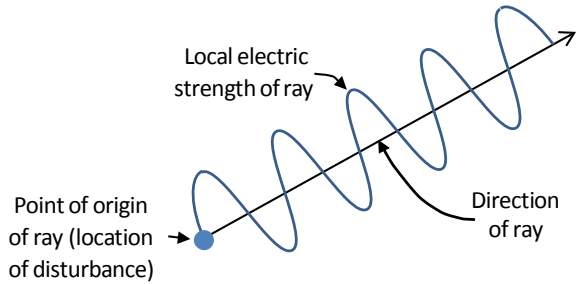

Figure 5. Anatomy of an individual ray.

Figure 6 illustrates equally well the geometry for both the infinite slit and the circular aperture diffraction problems. We consider a source point P' in the plane of the slit or circular aperture and a field point P lying on the screen. Then, referring to Fig. 6, the phase ,ϕ, of the ray when it arrives at screen point P will depend only on the wavelength λ of the light and the length of the line connecting source point P' with field point P; that is,

$$\phi = 2\pi \frac{z/\lambda}{\cos\vartheta}, \quad (5)$$

where z is the horizontal distance between the slit or aperture and the screen, and $\vartheta$ is the angle between the ray and the z-axis. The electric field strength of the ray at field point S' is then

$$E = E_0 e^{i\phi}. \quad (6)$$

Within an arbitrary constant the intensity distribution on the screen is given by

$$I(y') \propto E(y') \times E^*(y'), \quad (7)$$

where E(y') is the local electric field due to all of the rays incident to a given field point, and * denotes its complex conjugate. This calculation requires that the screen surface be divided into bins since it is unlikely that two rays will be incident at exactly the same point. Then the electric field strength in bin n is the algebraic sum of the contributions by the individual rays that are incident within the bin.

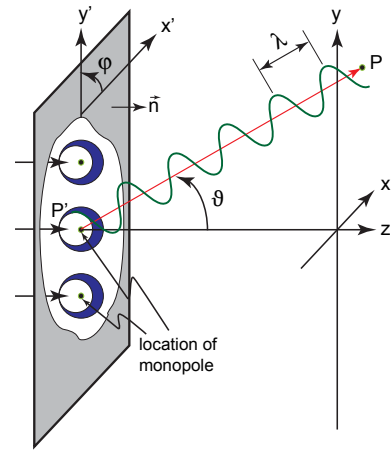

Figure 6. The MCRT diffraction model.

The validity of the Huygens principle illustrated in Fig. 4 may be questioned because of the presumed monopole nature of the disturbances forming the wavefront at each time step. At any instant, every point on the primary wavefront is envisioned as a continuous emitter of spherical secondary wavelets. Each wavelet radiated uniformly in all directions, in addition to generating an ongoing wave, thus, there would be also a reverse wave traveling back toward the source. No such a wave is found experimentally. Attempts to address this inconsistency when using the Huygens principle as the basis for formulating diffraction by apertures have led to the concept of the obliquity factor. The obliquity factor attributed to Kirchhoff has the form K($\vartheta$) = (1 + cos$\vartheta$)/2 at a given angle $\vartheta$ with respect to the aperture normal **n**. Spherical secondary wavelets with weighted direction have been shown in the Figure 6. [14] This has its maximum value, K(0) = 1, in the forward direction and disperses with the back wave, K($\pi$) = 0. Obliquity factors have been used with varying degrees of success in analytical treatments of diffraction; however, their possible role in MCRT models has been largely ignored. The traditional role of the obliquity factor is to properly model the variation of amplitude with angle $\vartheta$ for each refracted wavelet.[13] Taking into account the Huygens-Fresnel principle and the obliquity factor contribution, the optical rays are randomly generated and uniformly distributed in the single slit or the circular apertures. An equivalent approach, arguably more convenient to use in the MCRT environment, is to assign the same power to all refracted wavelets but to adjust their angular density distribution to account for obliquity. The random points are uniformly distributed in the single slit. For circular apertures, random points are homogeneously distributed over a unit disk in the form

$$\vartheta = \sin^{-1}\left[\sqrt{R_\vartheta}\right], \quad (8)$$

and

$$\varphi = 2\pi R_\varphi, \quad (9)$$

where $\vartheta$ is the zenith angle measured from the aperture surface normal, **n**, $\varphi$ is the azimuth angle measured in the plane of the aperture from an arbitrary reference, and lies in the plane of the aperture surface (Fig. 6), and $R_\vartheta$ and $R_\varphi$ are random numbers whose values are uniformly distributed between zero and unity. In the MCRT view of refraction, illustrated in Figure 6, the refraction event occurs when the rules of random points for single slit and circular apertures are applied, where the rays abruptly change directions. Following the Huygens-Fresnel principle, each ray is divided into a number of rays, called refractions per ray. The complex amplitude at the point P in Figure 6 is found by the superposition of waves or summing contributions from each point on the sphere of the primary wavelets.

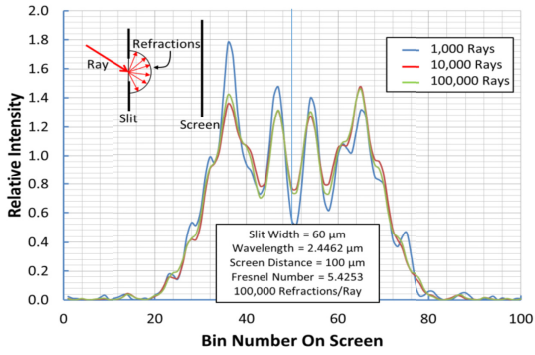

Figure 7. (Color online) Sensitivity of the MCRT-predicted fringe pattern to the number of rays traced.

## 6. RESULTS AND DISCUSSION

We define the Fresnel number

$$F \equiv a\sqrt{2/\lambda z}, \qquad (10)$$

where a is the slit width or aperture diameter and z is the distance from the aperture to the screen upon which fringes are formed. Two fringe patterns produced by an aperture at the same value of Fresnel number are known to be formally similar, regardless of the values of a, λ, and z. By convention, if F > 1.0, diffraction is considered to be in the Fresnel regime, and if F < 1.0 diffraction is considered to be in the Fraunhofer regime.

Figure 7 illustrates the fringe pattern predicted using the MCRT method for the case of a 60-μm slit at a wavelength of 2.4462 μm and a screen distance of 100 μm, corresponding to a Fresnel number of 5.4253. Results are shown for three values of the number of rays traced from randomly located positions y in the slit: one thousand, ten thousand, and one hundred thousand. The results for this case, which can be considered typical, verify that convergence is assured when 100,000 refractions per ray are launched from 100,000 randomly located points in the slit. A measure of the accuracy of the results obtained can be assessed by observing the departure from symmetry of the results about the center plane located at bin number 50.

Experience has shown that the MCRT results are also sensitive to the number of bins into which the wavelets are bundled on the screen. The pinhole aperture of the detector in Fig. 1 has a diameter of 2 μm, and the experimental results reported here are for slit widths and aperture diameters of 100 and 200 μm. Therefore, the MCRT bin size roughly corresponds to the measurement spatial resolution.

The diffraction fringe patterns vary strongly by varying the aperture-to-screen distance, z, from the near-field Fresnel region to the Fraunhofer regime. Figures 8a through 8f compare the diffraction fringe pattern profiles produced by a slit-type aperture for six values of Fresnel number ranging from 10.18 to 0.80. In this series of images, we vary the Fresnel number by varying the aperture-to-screen distance z for a value of slit width a = 200 μm and a fixed wavelength of λ = 351 nm. The profiles are taken at the half-length of a slit whose length is long compared to the slit width. The blue curves (bottom) are experimental results collected from our optical setup. The red curves (middle) are calculation using the Fresnel approach for the near-field Fresnel regime and the Fraunhofer approximation for the far-field region. The green curves (top) are the MCRT results described in the previous section. For this simulation, we have used 20,000 refractions per ray that are launched from 20,000 randomly located points in the 200-μm single slit. The theoretically calculated curves from the Fresnel approach as well as the Fraunhofer approximation are remarkably similar to the observed diffraction patterns. The computation time for the MCRT method for the setup is about 3 minutes. The time for Fresnel calculation using Fourier transform is significant shorter around one minute with the same computer hardware.

The MCRT method is a numerical analysis. The advantage of the numerical method is that there is virtually no limit to solve complex problems including geometrical symmetry, but this method produces numerical errors. In the following part, we provide a comparison between the MCRT method and the analytical analysis based on Fresnel/Fraunhofer approach with experimental results. The phase, ϕ, electric field, E, as well as the intensity, I, of an optical ray at screen point P will depend on the wavelength, λ, of the light, and the length of the line connecting source point P' with field point P. Thus, we extract the phase information, electrical field and intensity of refracted optical rays that launched from randomly located points in the aperture slits using the MCRT method.

We have obtained the diffraction fringe patterns by changing the size of the single-slit aperture. Figures 9a, 9b, and 9c compare the diffraction fringe pattern profiles produced by a slit-type aperture for three values of the Fresnel number ranging from 6.01 to 4.03. In this series of images, the slit width a = 100 μm and the wavelength λ = 351 nm. Comparison of Figs. 8b with Fig. 9a and Fig. 8c with Fig. 9b verifies the formal similarity of fringes corresponding to the same (approximately in these cases) value of Fresnel number. These experimental results (blue curves) are compared with MCRT (green curves) and analytical (red curves) simulations.

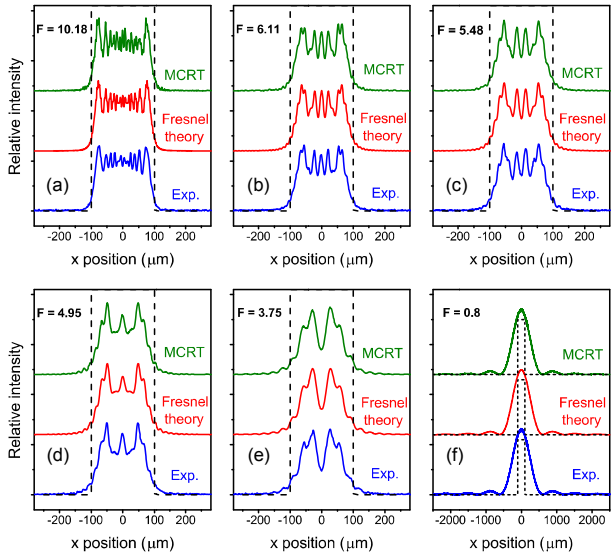

Figure 8. (Color online) Comparison of MCRT, theoretical, and measured diffraction fringes produced by a 200-μm slit illuminated by a 351-nm laser for a range of aperture-to-fringe distances.

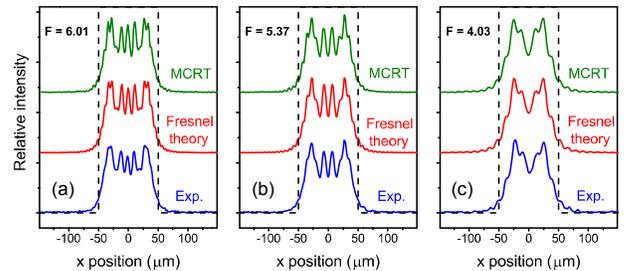

Figure 9. (Color online) Comparison of MCRT, theoretical, and measured diffraction fringes produced by a 100-μm slit illuminated by a 351-nm laser for a range of aperture-to-fringe distances.

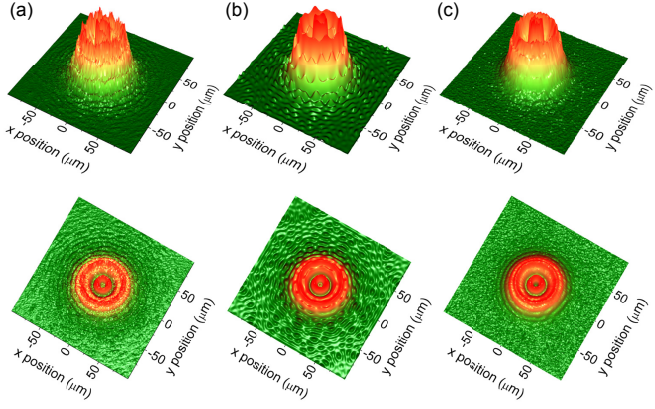

Figure 10. (Color online) Diffraction fringes computed (a) using the MCRT method, (b) using Fresnel theory, and (c) measured corresponding to normal illumination of a 100-μm diameter circular aperture by a 351-nm laser with a screen distance of z = 1.9 mm (F = 5.48).

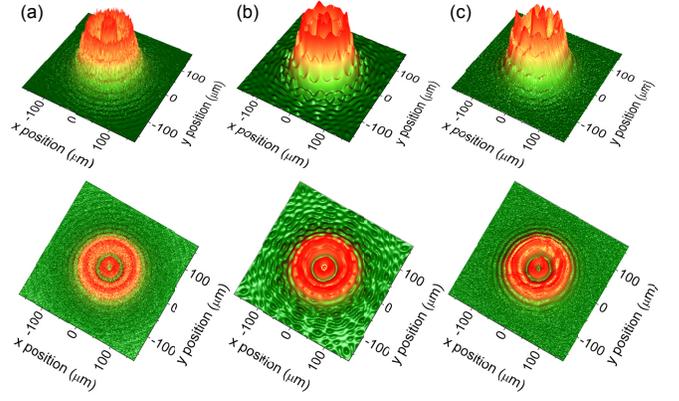

Figure 11. (Color online) Diffraction fringes computed (a) using the MCRT method, (b) using Fresnel theory, and (c) measured corresponding to normal illumination of a 200-μm diameter circular aperture by a 351-nm laser with a screen distance of z = 7.6 mm (F = 5.48).

We have measured diffraction fringes produced by circular apertures, and these results are compared with MCRT and analytical results (Figs 10 to 12). The fringes in all three figures correspond to normal illumination of the aperture by an expanded 351-nm laser beam. In Fig. 10, the aperture diameter is 100 μm with a screen distance of 1.9 mm, in Fig. 11 the diameter is 200 μm with a screen distance of 7.6 mm, and in Fig. 12 the diameter is 400 μm with a screen distance of 30.4 mm. The Fresnel number in all three figures is the same, F = 5.48; therefore, we expect fringe patterns to have identical shapes even though they cover different surface areas on the screen.

Intercomparison of the three figures reveals that the diffraction fringes are indeed formally similar even though the surface area they cover on the screen increases with increasing screen distance z, as expected. As a consequence of increasing surface area with increasing screen distance, the spatial resolution of the measured fringes increases going from Fig. 10 to Fig. 12. However, the sampling of the MCRT-based images decreases as a fixed number of rays traced is spread over a larger screen area. This leads to an increasing "fuzziness" of the MCRT-based images moving from Fig. 10 to Fig. 12. All three figures exhibit excellent agreement among the MCRT-based, analytical, and measured images.

A comparison between the experimental data and the theoretical models has been performed by a careful application of standard statistical analysis including the adjusted coefficient of determination, $R^2_{\text{adj}}$, the root-mean-square deviation, RMSD, and the reduced chi-square statistics, $\chi^2_\nu$. In statistics, the most common measure is the coefficient of determination, $R^2$, that gives information about the goodness of fit of a model. The $R^2$ coefficient of determination provides an estimation of how well observed results are replicated by the model, based on the proportion of total variation of theoretical values.[21]

$$R^2 = 1 - \frac{RSS}{TSS} = 1 - \frac{\sum_{i=0}^{N}(I_i^{\text{exp}}-I_i^{\text{mod}})^2}{\sum_{i=0}^{N}(I_i^{\text{exp}}-\bar{I})^2}, \qquad (11)$$

where RSS is the residual sum-of-squares, TSS is the total sum-of-squares, $I_i^{\text{exp}}$ is the $i$th observed value of N observations, $I_i^{\text{mod}}$ is the corresponding theoretical value and $\bar{I}$ is mean of the observed data. In general, the larger value of $R^2$, the better the agreement between experimental results and theoretical model. In the linear context, this measure is very intuitive as values between 0 and 1 provide a ready interpretation of of the degree to which the variance in the data is explained by the theoretical model. The value of $R^2$ will always increase when a new independent value is added. This is counter to the intuitive expectation that a theoretical model with more independent variables should provide a better fit. To compensate for the possible bias due to different number of parameters, we employ the adjusted coefficient of determination, $R^2_{\text{adj}}$:

$$R^2_{\text{adj}} = 1 - \frac{N-1}{N-p-1} \times (1-R^2), \qquad (12)$$

Table 1: Summary of a comparison between MCRT simulation and Fresnel/Fraunhofer approach with experimental results.

|  | Model | $R^2_{\text{adj}}$ | RMSD | $\chi^2_\nu$ |
|---|---|---|---|---|
| **Single slit (200 μm)** | | | | |
| F = 0.80 | MCRT | 0.99820 | 4.09 × 10⁻⁴ | 1.28 |
| | Fraunhofer approximation | 0.99756 | 4.79 × 10⁻⁴ | 1.54 |
| F = 3.75 | MCRT | 0.99611 | 2.51 × 10⁻⁴ | 0.48 |
| | Fresnel theory | 0.98654 | 4.39 × 10⁻⁴ | 0.36 |
| F = 4.95 | MCRT | 0.99531 | 3.01 × 10⁻⁴ | 0.46 |
| | Fresnel theory | 0.99439 | 3.29 × 10⁻⁴ | 0.43 |
| F = 5.48 | MCRT | 0.99278 | 3.66 × 10⁻⁴ | 1.88 |
| | Fresnel theory | 0.99017 | 4.24 × 10⁻⁴ | 0.68 |
| F = 6.11 | MCRT | 0.99019 | 4.29 × 10⁻⁴ | 1.91 |
| | Fresnel theory | 0.98680 | 4.97 × 10⁻⁴ | 0.91 |
| F = 10.18 | MCRT | 0.97186 | 7.72 × 10⁻⁴ | 1.96 |
| | Fresnel theory | 0.98230 | 6.15 × 10⁻⁴ | 1.34 |
| **Single slit (100 μm)** | | | | |
| F = 4.03 | MCRT | 0.98549 | 4.28 × 10⁻⁴ | 0.62 |
| | Fresnel theory | 0.99224 | 3.13 × 10⁻⁴ | 0.43 |
| F = 5.37 | MCRT | 0.99472 | 2.56 × 10⁻⁴ | 1.05 |
| | Fresnel theory | 0.99517 | 2.45 × 10⁻⁴ | 0.81 |
| F = 6.01 | MCRT | 0.98708 | 4.09 × 10⁻⁴ | 1.62 |
| | Fresnel theory | 0.98703 | 4.09 × 10⁻⁴ | 0.89 |

where $p$ is independent variables. $R^2_{\text{adj}}$ is always smaller than $R^2$. The independent variables are unknown parameters of our calculations, $p = 6$, including the slit width, the distance from the aperture to the screen, the wavelength, the number aperture points, the number of rays traced per aperture point, and the number of bins on the screen. The value of $p$ is much smaller than the number of observations with $N = 401$.

Using $R^2$ or adjusted $R^2_{\text{adj}}$ alone is not sufficient; it is also necessary to diagnose regression results by a residual analysis. We employ here the root-mean-square deviation (RMSD) or the root-mean square error (RMSE), to assess the quality of a regression. The RMSD is a frequently used measure of differences between values predicted by a model and observed data. The RMSD provides an aggregation of magnitudes of errors between predictions and observed values. Thus, the RMSD,

$$RMSD = \sqrt{\frac{\sum_{i=0}^{N}(I_i^{\text{exp}} - I_i^{\text{mod}})^2}{N}}, \quad (13)$$

is a measure of accuracy to compare forecasting errors of different models for a particular data set and not between datasets.

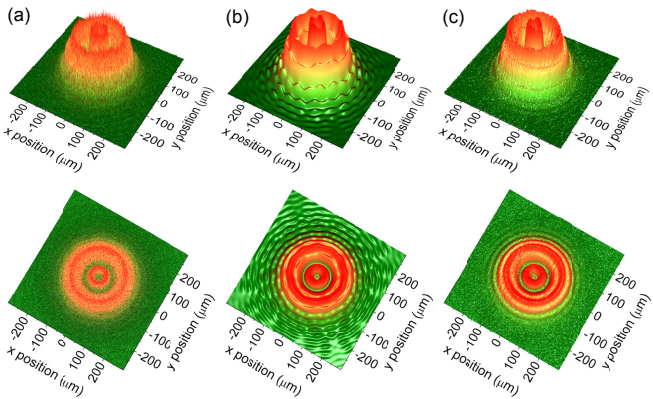

Figure 12. (Color online) Diffraction fringes computed (a) using the MCRT method, (b) using Fresnel theory, and (c) measured corresponding to normal illumination of a 400-μm diameter circular aperture by a 351-nm laser with a screen distance of z = 30.4 mm (F = 5.48).

The variance of a least-squares regression analysis is also characterized by the chi-square statistics, $\chi^2$ [36]:

$$\chi^2 = \sum_{i=0}^{N} \frac{(I_i^{\text{exp}} - I_i^{\text{mod}})^2}{\sigma_i^2}, \quad (14)$$

where $\sigma_i^2$ is the uncertainty in individual measurements, $I_i^{\text{exp}}$. We further define the reduced chi-square, $\chi^2_\nu$, as a useful measure by:

$$\chi^2_\nu = \frac{1}{N-p} \sum_{i=0}^{N} \frac{(I_i^{\text{exp}} - I_i^{\text{mod}})^2}{\sigma_i^2}, \quad (15)$$

where $\nu = N - p$ is the degrees of freedom. As a general rule, a value of reduced chi-square, $\chi^2_\nu \gg 1$, indicates a poor agreement between experimental results and the theoretical model. If the theoretical model is a good approximation, then the variances of both should be in good agreement, and the reduced chi-square should be approximately unity, $\chi^2_\nu \sim 1$. If the reduced chi-square is too small, $\chi^2_\nu \ll 1$, it may indicate that one has been too pessimistic about measurement errors.

We have performed least-squares regression analyses including the adjusted coefficient of determination, $R^2_{\text{adj}}$, the root-mean-square deviation, *RMSD*, and the reduced chi-square statistics, $\chi^2_\nu$, for the MCRT method and the analytical approach to model diffraction irradiation patterns with different distances from apertures to the observation. Table I provides calculation of these parameters for different values of Fresnel numbers and different sizes of single slits. The values of the adjusted coefficient of determination, $R^2_{\text{adj}}$, are near unity. This indicates an excellent agreement between observation and both the MCRT and Fresnel/Fraunhofer models for the diffraction patterns.

The RMSD is very small for both models. These values also indicate that the MCRT as well as analytical methods are excellent models for the diffraction patterns. The values of the reduced chi-square, $\chi^2_\nu$, are approximately unity. Therefore, statistical analysis confirms the qualitative observation that measured fringe data can be explained very well using both the MCRT method and the Fresnel/Fraunhofer approach. All three statistical methods confirm excellent agreement among the MCRT method, the standard analytical approach, and measured diffraction patterns.

## 7. CONCLUSIONS

The obliquity rule based on dipole radiation from each diffraction site produces excellent agreement between the MCRT diffraction model and corresponding experimental results when care is taken to assure that the screen binning precision in the MCRT model matches the experimental measurement precision.

**Acknowledgement.** The authors gratefully acknowledge the contributions of Dr. Kory J. Priestley, Senior Research Scientist in the Climate Science Branch at NASA's Langley Research Center.